\documentclass[prb,twocolumn,showpacs]{revtex4}
\usepackage{amssymb,amsmath,bm,epsfig,graphics,subfigure}
\pdfoutput=1
\begin{document}
\title{Band Symmetries and Singularities in Twisted Multilayer Graphene}
\author{E. J. Mele}
    \email{mele@physics.upenn.edu}
    \affiliation{Department of Physics and Astronomy University of Pennsylvania  Philadelphia PA 19104  \\}
\date{\today}

\begin{abstract}  The electronic spectra of rotationally faulted graphene
bilayers are calculated using a continuum formulation for small fault angles that identifies two distinct electronic states of
the coupled system. The low energy spectra of one state features a Fermi velocity reduction which ultimately leads to pairwise
annihilation and regeneration of its low energy Dirac nodes. The physics in the complementary state is controlled by pseudospin
selection rules that prevent a Fermi velocity renormalization and produce second generation symmetry-protected Dirac
singularities in the spectrum. These results are compared with previous theoretical analyses and with experimental data.
\end{abstract}


\maketitle The variation of the electronic properties of few layer graphenes (FLG's) with their layer stacking is receiving
increasing attention. FLG's represent a family of materials that bridge the pseudo relativistic properties of single layer
graphene with the more conventional semimetallic behavior of bulk graphite. The atomic registry of neighboring layers and
stacking sequence are structural parameters that determine their electronic properties  \cite{McCann,Ohta,guinea,MM,koshino}.
In twisted FLG's where the crystallographic axes of neighboring layers are misaligned by a rotation angle $\theta \neq n \pi/3$
the interlayer interactions produce remarkably rich physics that is being actively studied
\cite{Berger,LpD,Hass2,Latil,Shallcross,Sprinkle,miller,GTdL,MeleRC,Hicks,Bistritzer,LiAndrei,Li,KindermannFirst,Luican,deGail,choi}.

This paper presents a continuum theory of the low energy electronic physics in twisted bilayer graphenes for small rotation
angles, as illustrated in Fig. 1. Our approach reveals the existence of {\it two distinct} electronic states in this system
that present quite different electronic properties. The behavior of one state is identified with the situation described by a
frequently adopted continuum formulation of this problem \cite{LpD,Bistritzer}: the interlayer coupling renormalizes the Fermi
velocities of the individual layers and hybridizes their Dirac cones in the spectral region where they merge. In the
complementary state we find that the Fermi velocity renormalization is nearly completely prevented by a pseudospin selection
rule and the interlayer hybridization inherits a novel momentum space geometry producing a set of {\it second generation} Dirac
singularities. The behavior in this latter family agrees well with properties experimentally observed for rotationally faulted
FLG's thermally grown on SiC ${\rm (000 \bar 1)}$ \cite{Hass2,Sprinkle,Hicks} suggesting that this physics is realized in this
form of FLG. We briefly discuss the relation of our new results to prior theoretical and to experimental studies of these
systems.

The physics described below is identified by consideration of the effects of the {\it lattice symmetry} on the low energy
electronic physics. We show that the geometrical structure of the low spectrum is determined by a symmetry-allowed threefold
anisotropy in the interlayer coupling amplitudes which, though absent from conventional two-center tight binding models, occur
in empirical models of interlayer coupling in graphite. We find that the {\it sign} of this anisotropy distinguishes two quite
different electronic states of this system.

\begin{figure}
\begin{center}
\includegraphics[angle=0,width=\columnwidth,bb=0 0 360 400]{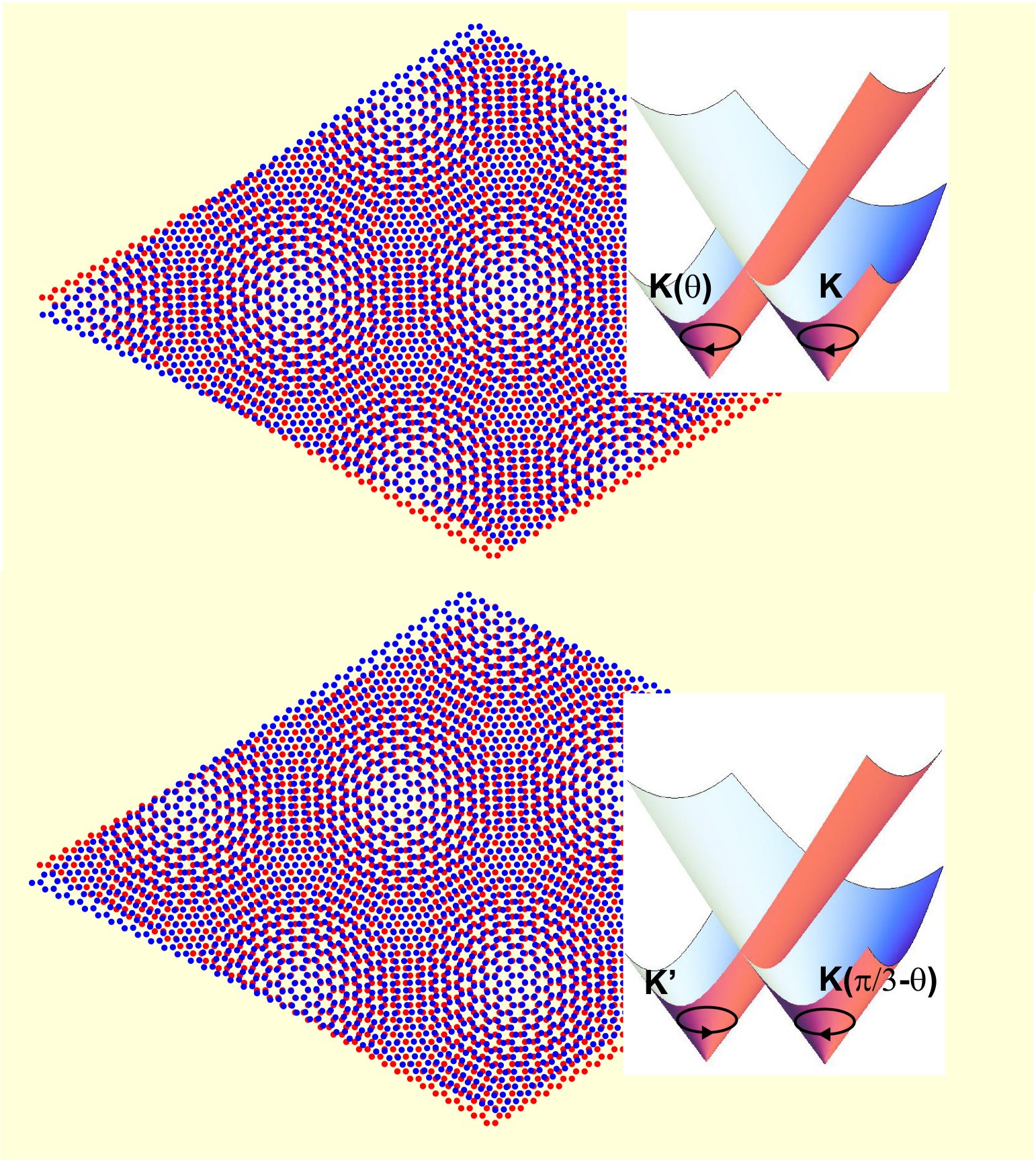}
\end{center}
\caption{\label{twolattices} Lattice structures of two twisted graphene bilayers rotated away from $AA$ stacking by angles
$\theta = 3.89^\circ$ (top) and $\theta = 56.11^\circ$ (bottom). The insets show schematically the dispersions of two nearby
Dirac cones in these structures in the absence of their interlayer coupling.}
\end{figure}

The coupling between the two sublattices in the two layers can be represented by a (position dependent) $2 \times 2$ matrix
operator $\hat T_{12}(\vec r)$. As shown in Figure 2, for small angle faults the registry between layers in the unit cell
evolves smoothly from regions locally characterized by $AB$ (region $\alpha$), $BA$ ($\beta$) and $AA$ ($\gamma$). The
smoothest possible supercell-periodic matrix-valued expression for $\hat T(\vec r)$ is given by the expansion
\begin{eqnarray}
\hat T_{12}(\vec r) = \hat t_0 + \sum_{n=1}^6 \, e^{i \vec {\cal G}_n \cdot \vec r} \hat t_n
\end{eqnarray}
with constant matrix coefficients $\hat t_n$ and where $\vec {\cal G}_n$ are the six elements of the first star of reciprocal
lattice vectors dual to the superlattice translations $\vec T_1$ and $\vec T_2$. The matrix coefficients $\hat t_n \, (n=1,6)$
can be determined from the couplings in the locally registered regions;  for example in the geometry of Figure 2 the even
elements of the first star have coefficients
\begin{eqnarray}
\hat t_{n \, {\rm even}} = t_{\cal G} \left(%
\begin{array}{cc}
  e^{-i \vec {\cal G}_n \cdot \vec r_\gamma} & e^{-i \vec {\cal G}_n \cdot \vec r_\alpha} \\
  e^{-i \vec {\cal G}_n \cdot \vec r_\beta} & e^{-i \vec {\cal G}_n \cdot \vec r_\gamma} \\
\end{array}%
\right) =  t_{\cal G} \left(%
\begin{array}{cc}
  z & 1 \\
  z^* & z \\
\end{array}%
\right)
\end{eqnarray}
where $z=e^{2 \pi i/3}$ and the coefficients for the odd elements are $t_{n \, {\rm odd}} = t_{n \, {\rm even}}^*$. The
constant matrix has the form
\begin{eqnarray}
\hat t_0 = \left(%
\begin{array}{cc}
  c_{aa} & c_{ab} \\
  c_{ba} & c_{bb} \\
\end{array}%
\right)
\end{eqnarray}
with real coefficients satisfying $c_{aa} = c_{bb}$ and $c_{ab} = c_{ba}$. The interlayer operator of Eqns. (2) and (3) is thus
parameterized by three real constants $t_{\cal G}$, $c_{aa}$ and $c_{ab}$. We choose these coefficients so that the interlayer
matrix $\hat T(\vec r_\alpha)$ matches the Slonczewski-Weiss-McClure (SWMcC) interlayer parameters $\gamma_1$, $\gamma_3$ and
$\gamma_4$ for Bernal stacked graphite  shown in the inset of Figure 2  \cite{millie} with the results in Table I. We note that
the $\gamma_3$ parameter (hopping between unaligned sublattices in the two layers) is comparable to $\gamma_1$ and that the
$\gamma_4$ parameter (hopping between aligned and unaligned sublattice sites) is relatively weak.

\begin{figure}
\begin{center}
\includegraphics[angle=0,width=\columnwidth,bb=0 0 2800 2100]{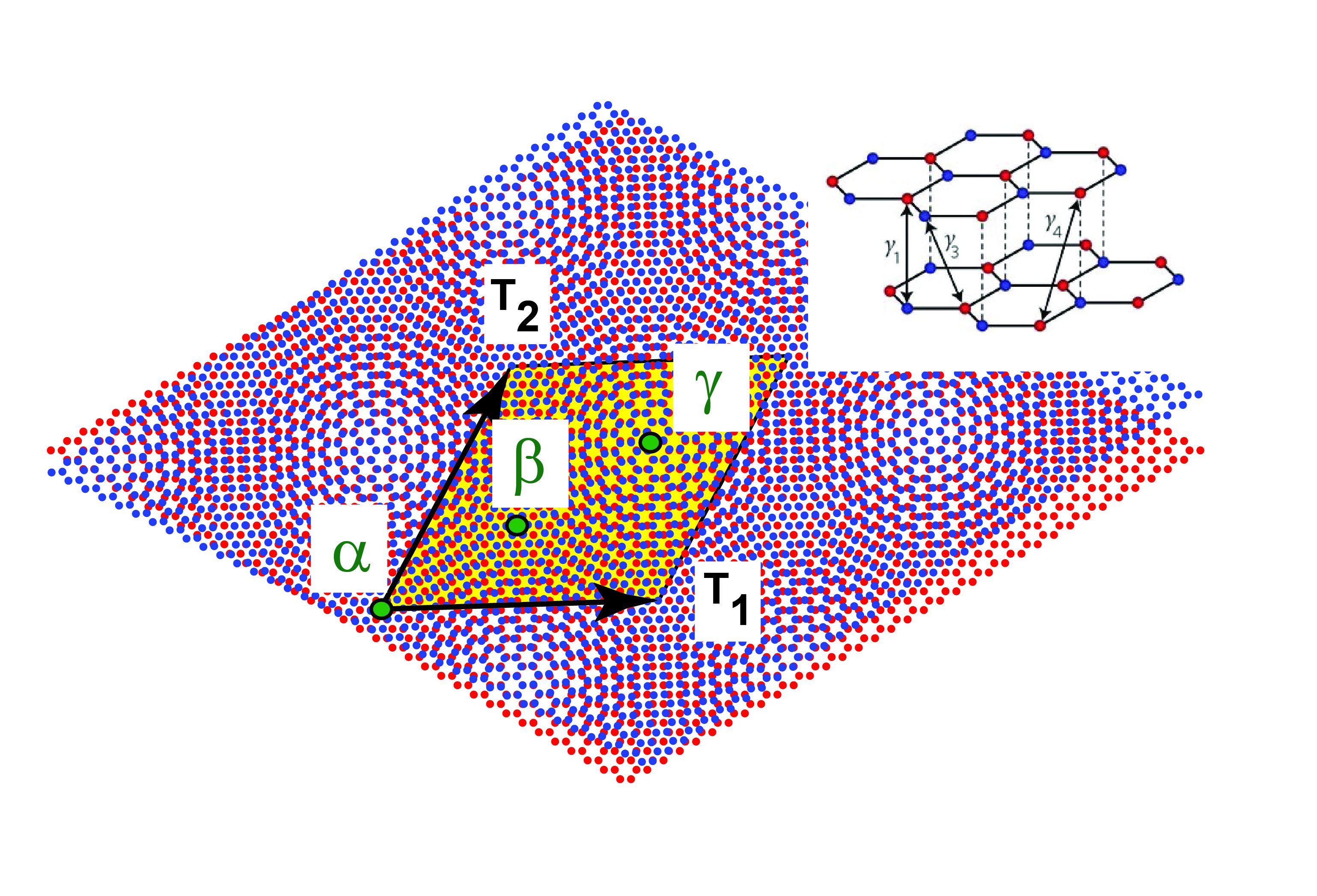}
\end{center}
\caption{\label{label}Lattice structure for a segment of twisted bilayer at rotation angle $3.89^\circ$, with superlattice
translation vectors $T_1$ and $T_2$. The points labelled $\alpha$, $\beta$ and $\gamma$ are high symmetry points in the unit
cell. The inset \cite{freitag} illustrates three hopping processes in the interlayer Hamiltonian. }
\end{figure}
\begin{table}
  \centering
\begin{tabular}{c c c c} % centered columns (4 columns)
\hline\hline %inserts double horizontal lines
Coefficient & Parameterization    & I & II  \\ [0.5ex] % inserts table
%heading
\hline % inserts single horizontal line
$t_{\cal G}$ & $(\gamma_1 - \gamma_3)/9$ & 43.3  & 8.3  \\ % inserting body of the table
$c_{aa}$ & $\gamma_4 + (\gamma_1 - \gamma_3)/3$ & 130.0 & 69.0  \\
$c_{ab}$ & $(\gamma_1 + 2 \gamma_3)/3$ & 130.0  & 340.0  \\ [1ex] % [1ex] adds vertical space
\hline\hline %inserts single line
\end{tabular}
\label{table:nonlin} % is used to refer this table in the text
  \caption{Fourier coefficients (meV units) for the  interlayer hopping operator Eqn. 2, fitted to the
  Slonczewski Weiss McClure parameterization for Bernal stacked layers. Model I:
  $\gamma_1 = 390$ meV, $\gamma_3=\gamma_4=0$. Model II: $\gamma_1 = 390$ meV, $\gamma_3$ = 315 meV and $\gamma_4 = 44$ meV.}\label{SWMcC}
\end{table}

The conventional continuum description of twisted bilayer graphene \cite{LpD,Bistritzer} can be derived from the constant
matrix $\hat t_0$. The low energy Hamiltonian is a long wavelength expansion around the zone corner points in each layer; in
this Dirac basis the matrix elements in  Eqn. 1 acquire the phases $\exp(i (\vec G' \cdot \vec \tau_i' - \vec G \cdot \vec
\tau_j))$ where $\vec G(\vec G')$ are reciprocal lattice vectors in the two separate layers and $\vec \tau_j (\vec \tau_i')$
are sublattice positions. Boosts by a triad of $(\vec G, \vec G')$ pairs translate the Hamiltonian to three pairs of zone
corner points  that are separated by $\Delta \vec K$ and its $\pm 2 \pi/3$-rotated counterparts. This generates three possible
constant coupling matrices indexed by the momentum differences $\Delta \vec K_i$. With a conventional choice of origin
\cite{LpD,Bistritzer} where the  $A$ sublattice site of one layer is aligned with the $B$ sublattice of the other, the matrices
are
\begin{eqnarray}
\hat T_1 &=& \left(%
\begin{array}{cc}
  1 & 0 \\
  0 & 1 \\
\end{array}%
\right) \hat t_0 \left(%
\begin{array}{cc}
  1 & 0 \\
  0 & 1 \\
\end{array}%
\right) = \left(%
\begin{array}{cc}
  c_{aa} & c_{ab} \\
  c_{ba} & c_{bb} \\
\end{array}%
\right) \nonumber\\
\hat T_2 &=& \left(%
\begin{array}{cc}
  1 & 0 \\
  0 & z \\
\end{array}%
\right) \hat t_0 \left(%
\begin{array}{cc}
  z & 0 \\
  0 & 1 \\
\end{array}%
\right) = \left(%
\begin{array}{cc}
  zc_{aa} & c_{ab} \\
  z^*c_{ba} & zc_{bb} \\
\end{array}%
\right) \nonumber\\
\hat T_3 &=& \left(%
\begin{array}{cc}
  1 & 0 \\
  0 & z^* \\
\end{array}%
\right) \hat t_0 \left(%
\begin{array}{cc}
  z^* & 0 \\
  0 & 1 \\
\end{array}%
\right) = \left(%
\begin{array}{cc}
  z^*c_{aa} & c_{ab} \\
  zc_{ba} & z^*c_{bb} \\
\end{array}%
\right) \nonumber\\
\end{eqnarray}
In one of these  valleys the Hamiltonian for the coupled bilayer with a momentum offset $\Delta \vec K$ is
\begin{eqnarray}
H = \left(%
\begin{array}{c  c}
    \hbar v_F \sigma \cdot (-i \nabla) & \hat T_1^\dag \\
  \hat T_1 & \hbar v_F \sigma_\theta \cdot (-i \nabla - \Delta \vec K)\\
\end{array}%
\right)
\end{eqnarray}
where $\sigma_\theta$ are Pauli matrices resolved along the axes of the $\theta$-rotated layer. The problem can be written
dimensionless form by scaling all momenta by the offset $|\Delta \vec K|$ and energies by $E_\theta = \hbar v_F |\Delta \vec
K|$. The scaled coupling coefficients are $\tilde c = c/E_\theta = 3ac/(8 \pi \hbar v_F \sin(\theta/2))$ (where $a$ is the
single layer graphene lattice constant) which increase with decreasing rotation angle.

Model I (Table I) is an isotropic interlayer model with $\gamma_3 = \gamma_4 = 0$.  For an isotropic coupling model
$c_{aa}=c_{bb}=w$ and the interlayer matrices are
\begin{eqnarray}
\hat T_1 = w \left(%
\begin{array}{cc}
  1 & 1 \\
  1 & 1 \\
\end{array}%
\right); \, \hat T_2 = w \left(%
\begin{array}{cc}
  z & 1 \\
  z^* & z \\
\end{array}%
\right); \,
\hat T_3 = w \left(%
\begin{array}{cc}
  z^* & 1 \\
  z & z^* \\
\end{array}%
\right)
\end{eqnarray}
with $w=130 \, {\rm meV}$.  The form of these matrices and their prefactor agree with the estimates ($w \approx 110 \, {\rm
meV}$) obtained from tight binding calculations \cite{LpD,Bistritzer}. Our construction shows that these terms project the
$q=0$ term of the interlayer potential into the Dirac $K$-point (pseudospin) basis thereby coupling the electronic states in
the two layers with identical crystal momenta. Since only the $q=0$ term in the coupling is retained it does not depend on a
relative lateral translation of the two layers, in agreement with earlier work \cite{Bistritzer} and physically reasonable
since for small twist angle a rigid layer translation produces insignificant changes to the Moire superlattice. Thus Model I
reproduces the existing continuum theoretic phenomenology of the coupled system, and the calculation leading to Eqn. 6 provides
an alternate (and compact) derivation of the effective Hamiltonian used in these earlier studies \cite{LpD,Bistritzer}. The
left panel of Fig. 3 shows the bilayer spectra computed in this model which shows the expected ($\theta$-dependent) reduction
of the Dirac cone velocities and a hybridization of the two branches in the spectral region where they merge.
\begin{figure}
\begin{center}
\includegraphics[angle=0,width=\columnwidth,bb=0 0 1000 800]{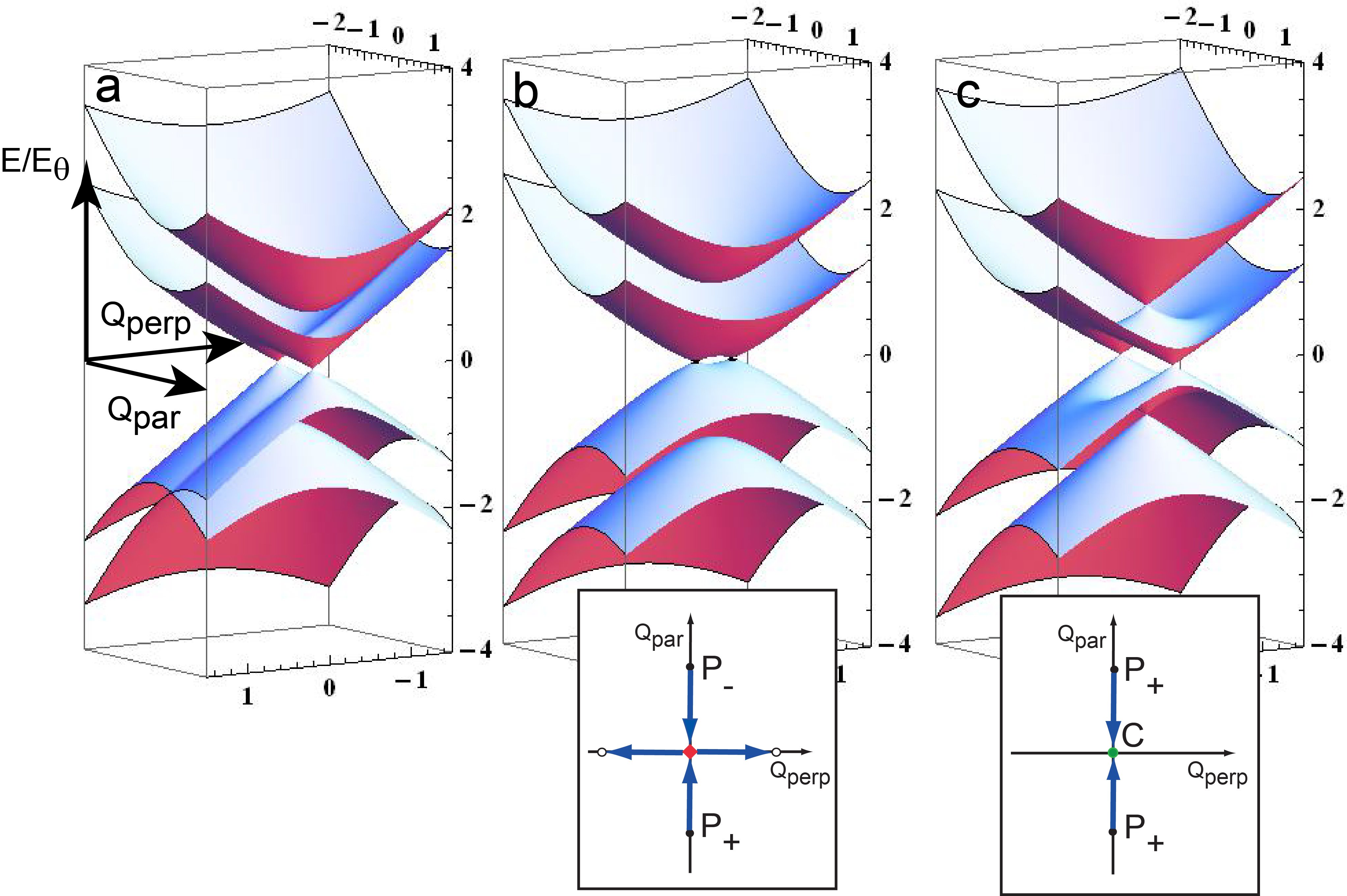}
\end{center}
\caption{\label{spectra} Electronic spectra for twisted bilayers using the interaction parameters (a): $\hat t_0 = \tilde c
(\texttt{I}+ \sigma_x), \, \tilde c = 0.21$, (b): $\hat t_0 = \tilde c \sigma_x, \tilde c = 0.55 $  and (c): $\hat t_0 = \tilde
c \texttt{I}, \tilde c = 0.55$. $Q_{\rm par}$ and $Q_{\rm perp}$ are momenta in units of the offset $|\Delta \vec K|$ and the
ordinate is the scaled energy $E/E_\theta = E/(\hbar v_F \Delta K)$. In (b) Dirac cones with opposite helicity are coupled, in
(c) Dirac cones with the same helicity are coupled. The insets give the locations of singular points in the spectra describing
the annihilation and regeneration of Dirac nodes (red diamond) in the compensated case (b) and the appearance of a singular
point of degeneracy (C) for the uncompensated case (c). The point C is a second generation Dirac point singularity in the
coupled spectrum.}
\end{figure}

We now consider a refinement of the interlayer coupling matrices using the parameterization of Model II.  The salient
properties of the SWMcC parameterization are the introduction of the interlayer amplitudes $\gamma_3$ and $\gamma_4$ with
$\gamma_3$ comparable to $\gamma_1$ and $\gamma_4$ significantly smaller. Note that $\gamma_3$ and $\gamma_4$ represent
interlayer hopping processes at the same range but in different directions with respect to the layer crystallographic axes. The
asymmetry between $\gamma_3$ and $\gamma_4$ thus reflects an intrinsic threefold lattice anisotropy in the interlayer
amplitudes which, though symmetry-allowed, does not occur in the isotropic two center tight binding approximation.
Significantly, these additional terms {\it break the symmetry} between the pseudospin-diagonal and off diagonal terms in $\hat
t_0$ (Table I) so that the coupling matrix is dominated by its off diagonal amplitudes. An instructive limit considers $\hat
t_0 \propto \sigma_x$ for which the Fig. 3(b) shows the spectrum calculated for a $\theta = 3.89^\circ$ rotation away from
Bernal stacking. Here the two Dirac cones have {\it merged} at low energy producing two composite low energy singular points.
Note that the linear low energy dispersion is replaced by an approximately quadratic form near the center of symmetry of these
spectra and that the momentum offset between the singular points in the spectrum is along the $Q_{\rm perp}$ axis, i.e.
$\pi/2$-rotated with respect to the original Dirac cone offset $\Delta \vec K$.

These spectral changes reflect the proximity to a  critical point that occurs at $\tilde c =1/2$ in this theory.  This can be
understood by considering a single layer sublattice exchange operation implemented by the gauge transformation
\begin{eqnarray}
\tilde H &=& \left(%
\begin{array}{cc}
   \texttt{I} & 0 \\
  0 & \hat \sigma_x \\
\end{array}%
\right) \left(%
\begin{array}{cc}
  \hat H_K(\vec q) & \tilde c  \hat \sigma_x \\
  \tilde c \, \hat \sigma_x & \hat H_K(\vec q - \Delta \vec K) \\
\end{array}%
\right)  \left(%
\begin{array}{cc}
 \texttt{ I} & 0 \\
  0 & \hat \sigma_x \\
\end{array}%
\right) \nonumber\\
&=& \left(%
\begin{array}{c|c}
  \hat H_K(\vec q) & \tilde c  \,  \texttt{I}\\
  \hline
  \tilde c \texttt{I} & \hat \sigma_x \cdot \hat H_K(\vec q - \Delta \vec K) \cdot \hat \sigma_x \\
\end{array}%
\right)
\end{eqnarray}
demonstrating that this system has a scalar coupling Dirac cones with compensating helicities (Berry's phase $\pm \pi$).
Increasing the control parameter $\tilde c$ (by decreasing $\theta$) draws the nodes together until they become coincident at a
critical coupling strength $\tilde c = 1/2$ and annihilate (Fig. 3b inset).  For $\tilde c > 1/2$ new singularities emerge at
$E=0$ separated by $\Delta \vec Q$ directed {\it perpendicular} to the original offset $\Delta \vec K$. Using the parameters
listed in Table II, $\tilde c \,(\theta =3.89^\circ) = 0.55$, i.e. just on the strong coupling side of this transition. The
residual curvature in the low energy spectrum and the associated reorientation of $\Delta \vec Q$ are both clearly evident in
Fig. 3b. It is noteworthy that the momentum separation between the zero energy contact points is generally {\it not} determined
purely geometrically by the rotation angle as is generally assumed, but instead is modified by the interlayer coupling. This
occurs because the interactions between layers produces an effective gauge field seen within each layer that shifts the
momentum of its zero energy states. The $\pi/2$ rotation of the momentum offset that bridges the contact points on the strong
coupling side of the transition is a striking consequence of this gauge coupling.

Reversing the {\it sign} of the threefold anisotropy in the interlayer matrix $\hat t_0$ produces a distinct electronic state.
The complementary behavior is understood  by considering the limit $\hat t_0 \propto \texttt{I}$ which describes the coupling
of Dirac cones with the {\it same} helicity, preventing annihilation of the Dirac nodes and leading to a qualitatively
different geometry in the bilayer spectrum  (Fig. 3c). The dispersing bands from the uncoupled cones are  degenerate everywhere
along the line that bisects $\Delta \vec K$. However, along the line that connects the Dirac nodes the pseudospins are {\it
orthogonal} and the interlayer coupling is symmetry-forbidden, turning on linearly as a function of the transverse momentum
$Q_{\rm perp}$. Thus the coupled system retains a twofold point degeneracy midway between the displaced Dirac nodes
\cite{offset}. The cancellation of the interlayer coupling at this critical point is the bilayer analog of the ``absence of
backscattering" due to the $\pi$ Berry's phase in single layer graphene. In the vicinity of this critical point interlayer
coupling is allowed and proportional to the transverse momentum. Thus this system exhibits {\it second generation Dirac
singularities} in its coupled layer spectrum as shown in Fig. 3c: hybridization of the two layers is symmetry forbidden at a
discrete critical crossing point. We refer to this complementary state as the uncompensated bilayer state.

The relative helicity of the two Dirac cones also controls the renormalization of their Fermi velocities, further
distinguishing these two states. For Dirac cones of opposite helicities, perturbation theory on the Hamiltonian in Eqn. 7 for
small $\tilde c$ modifies the velocity operators
\begin{eqnarray}
\hat v_+ = v_F \sigma_+ \rightarrow v_F (1 - \tilde c^2) \sigma_+ \nonumber\\
\hat v_- = v_F \sigma_- \rightarrow v_F (1 - \tilde c^2) \sigma_-
\end{eqnarray}
which symmetrically reduces both $v_x$ and $v_y$; summation over the triad of offset momenta $\Delta \vec K_i$ yields the
renormalized velocity $v_F^* = v_F (1 - 9 \tilde c^2)$ exactly as found in earlier work \cite{LpD,Bistritzer}. By contrast for
coupling between nearby cones of the same helicity perturbation theory yields
\begin{eqnarray}
\hat v_+ = v_F \sigma_+ \rightarrow v_F ( \sigma_+ - \tilde c^2 \sigma_-) \nonumber\\
\hat v_- = v_F \sigma_- \rightarrow v_F ( \sigma_- - \tilde c^2 \sigma_+)
\end{eqnarray}
so that in this class the corrections to the velocity  are weaker, $\propto \tilde c^4$. Moreover they have a {\it twofold}
$\cos(2 \phi)$ anisotropy so they vanish by symmetry after summing over the threefold symmetric triad of $\Delta \vec K_i$.
Thus the Fermi velocity is unchanged by the interlayer coupling in this class of bilayers. Physically this can be understood by
observing that the bands dispersing through the Dirac nodes are connected smoothly to the second generation points of degneracy
at $\Delta \vec K/2$.

The distinction between the compensated and uncompensated states in the small angle limit reflects a lattice-scale property
that determines the matrix structure of the long wavelength coupling in Eqn. 1.  This should be distinguished from the
different mechanism by which sublattice exchange symmetry determines the direct coupling between the Dirac nodes \cite{MeleRC}.
The latter requires {\it finite momentum} umklapp interlayer hopping processes which, though significant for low order rational
commensurate rotations, are negligible in the small angle limit considered here. For example, note that sublattice exchange
``even" and ``odd" commensurations are related by a rigid sublattice translation of one layer at a {\it fixed} rotation angle.
In the small angle regime this translation simply permutes regions of the bilayer that are locally in $AB$, $BA$ and $AA$
registry as shown in Fig. 1, but it does not change $\hat t_0$ which determines the spectrum. Thus sublattice exchange ``even"
and ``odd" structures become indistinguishable in the small angle limit. Note also that bilayers at rotation angles $\theta$
and $\bar \theta = \pi/3 - \theta$ are commensuration pairs that can be distinguished by their sublattice exchange parity
\cite{MeleRC}. Even and odd parity commensurations are, respectively, inflated generalizations of the primitive $AA$  and $AB$
stacked bilayers. This symmetry ultimately determines the valley structure of the interlayer amplitudes that directly couple
the Dirac nodes of neighboring layers.  This interlayer umklapp coupling derives from the finite momentum terms in the
interlayer Hamilonian in contrast to the $q=0$ terms that control the physics for small angle rotations.

The spectra for these two classes are ultimately determined by the pseudospin asymmetry in $\hat t_0$. The conventional SWMcC
model selects the class that couples cones with compensating helicities. In this model the spectral transition illustrated in
Figure 3 occurs for rotation angles near $4 ^\circ$, i.e. in a range that is frequently studied experimentally
\cite{LiAndrei,Li}. The physics of the uncompensated class occurs for $c_{aa} > c_{ab}$ which requires $\gamma_4 > \gamma_3$.
Although this is excluded by the conventional SWMcC parameterization it is important to note that this parameterization is
designed to fit data for Bernal stacking, and it likely does not properly represent the matrix structure of the coupling in
$AA$ registered regions. In particular using the parameterization of Table I, the spatial dependence of Eqn. 1 shows that
strong interlayer coupling in $AA$ stacked regions  requires $\gamma_4 > \gamma_3$. Microscopically this originates from
interlayer tunneling processes along the edges of eclipsed hexagons in the aligned $AA$ structure, a motif which does not
appear at all for Bernal stacking. In the spirit of the SWMcC theory it is therefore appropriate to retain $\gamma_3$ and
$\gamma_4$ as parameters which can be determined from the experimentally observed properties of twisted graphenes.

In fact the phenomenology of the uncompensated class provides a striking explanation for many of the puzzling observed spectral
properties for rotationally faulted graphenes thermally grown on ${\rm SiC(000 \bar 1)}$ \cite{miller,Sprinkle,Hicks}. Landau
level spectroscopy shows a negligible renormalization of the Fermi velocity in these structures \cite{miller}  and furthermore
angle resolved photoemission finds no evidence for a hybridization-induced avoided crossing of the intersecting Dirac cones,
despite a careful search \cite{Hicks}. This is completely consistent with the existence of a node in the interlayer coupling at
the midpoint between offset Dirac cones characteristic of the uncompensated class. This assignment can be confirmed
definitively by measurements of the quasiparticle dispersion along an azimuth passing through the midpoint between the
displaced Dirac cones, but {\it perpendicular} to $\Delta \vec K$;  these should show a band splitting linear in the transverse
momentum around the point of degeneracy. Alternatively, if these bilayers exist in the compensated class, photoemission should
be able to detect the annihilation and re-emergence of their singular contact points along with the band curvature in their
spectra in the crossover regime as illustrated in Fig. 3(b).

By contrast, experiments on rotationally faulted CVD-grown graphenes have observed phenomena that have been associated with the
spectral properties of the compensated class \cite{Luican,Li}. Features due to the van Hove singularities arising from an the
avoided crossing of hybridized Dirac cones \cite{Li} and a $\theta$-dependent low energy velocity renormalization have both
been reported \cite{Luican}. These features are at least qualitatively consistent with the predicted behavior of the
compensated class and have been analyzed within a theoretical model representative of this class \cite{LpD}. We note that these
measurements study samples at small rotation angle where the proximity to the merger of the Dirac singularities (Fig. 3) should
be manifest in these data though their effects have not yet been considered in the analysis. It is interesting that these
samples exhibit a large periodic height modulation $\approx 1 {\rm \AA}$ in the superlattice unit cell peaked in the
$AA$-registered zones \cite{height}. It is tempting to speculate that these CVD samples are grown as rippled structures that
partially delaminate in these regions thereby locally weakening their contribution to the $q=0$ coupling coefficients. In this
scenario the strongly coupled regions would maintain Bernal registry as described by the conventional SWMcC parameterization
and identify these samples as members of the compensated bilayer family.

The distinction between the two complementary states is controlled by an important three-fold anisotropy in the interlayer
tunneling amplitudes. This physics is not captured by an isotropic two-center tight binding theory, which inevitably leads one
to the coupling model in Eqn. 6 which happens to occur at a crossover between two rather different electronic models for the
system. The effects of the threefold anisotropy are accessible in density functional calculations of these structures, but for
practical reasons these have been restricted to short period superlattices which do not address the small angle regime where
the continuum theory is most appropriate. For short period commensurate structures, the Fermi velocities found in these
calculations are consistent with the values for single layer graphene. This could arise from the small value of $\tilde c$ in
the large angle regime, the intrinsic behavior of the uncompensated class or an interlayer mass term which is important for
short period superlattices \cite{MeleRC}.

I thank P. First, C. Kane, M. Kindermann, S. Zaheer and F. Zhang for their comments on the manuscript E. Andrei for
communication of unpublished data. This work was supported by the Department of Energy, Office of Basic Energy Sciences under
contract DE-FG02-ER45118.


\begin{thebibliography}{11}
\expandafter\ifx\csname natexlab\endcsname\relax\def\natexlab#1{#1}\fi
\expandafter\ifx\csname bibnamefont\endcsname\relax
  \def\bibnamefont#1{#1}\fi
\expandafter\ifx\csname bibfnamefont\endcsname\relax
  \def\bibfnamefont#1{#1}\fi
\expandafter\ifx\csname citenamefont\endcsname\relax
  \def\citenamefont#1{#1}\fi
\expandafter\ifx\csname url\endcsname\relax
  \def\url#1{\texttt{#1}}\fi
\expandafter\ifx\csname urlprefix\endcsname\relax\def\urlprefix{URL }\fi
\providecommand{\bibinfo}[2]{#2}
\providecommand{\eprint}[2][]{\url{#2}}

\bibitem{McCann} E. McCann and V.I. Fal'ko, Phys. Rev. Lett {\bf 96}, 086805 (2006).

\bibitem{Ohta}T. Ohta {\it et al.},  Science {\bf 313}, 951 (2006).

\bibitem{guinea} F. Guinea, A.H. Castro Neto, and N.M.R. Perez, Phys. Rev. B {\bf 73}, 245426 (2006).

\bibitem{MM} H. Min, A.H. MacDonald, Phys. Rev. B {\bf 77}, 155416 (2008).

\bibitem{koshino} M. Koshino and E. McCann, Phys. Rev. B {\bf 80}, 165409 (2009).

\bibitem{Berger} C. Berger {\it et al.}, Science {\bf 312}, 1191 (2006).

\bibitem{Latil} S. Latil {\it et al.},  Phys. Rev. B {\bf 76}, 201402(R) (2007).

\bibitem{LpD} J.M.B. Lopes dos Santos, N.M.R. Peres, A.H. Castro Neto,  Phys. Rev. Lett. {\bf 99}, 256802 (2007).

\bibitem{Hass2}J. Hass {\it et al.},   Phys. Rev. Lett. {\bf 100}, 125504 (2008).

\bibitem{Shallcross} S. Shallcross, S. Sharma  and  O.A. Pankratov, Phys. Rev. Lett.{\bf 101}, 056803 (2008).

\bibitem{Sprinkle} M. Sprinkle {\it et al.} Phys. Rev. Lett. {\bf 103}, 226803 (2009).

\bibitem{miller} D.M. Miller {\it et al.} Science {\bf 324}, 9242 (2009).

\bibitem{GTdL} G. T. de Laissardi$\grave{\rm e}$re {\it et al.} Nano Lett.\ {\bf 10}, 804 (2010).

\bibitem{MeleRC} E. J. Mele, Physical Review B {\bf 81}, 161405 (2010).

\bibitem{Hicks} J. Hicks {\it et al.}, Phys. Rev. B {\bf 83}, 205403 (2011).

\bibitem{Bistritzer} R.Bistritzer and A.H. MacDonald,  Proc. Nat. Acad. Sci. {\bf
108},12233 (2011).

\bibitem{LiAndrei} G. Li, A. Luican and  E.Y. Andrei,  Phys. Rev. Lett. {\bf 102}, 176804 (2009).

\bibitem{Li} G. Li {\it et al.} Nature Physics {\bf 6}, 109 (2010).

\bibitem{KindermannFirst} M. Kindermann and P.N. First, Physical Review B {\bf 83}, 045425 (2010).

\bibitem{Luican} A. Luican {\it et al.} Phys. Rev. Lett. {\bf 106}, 126802 (2011).

\bibitem{deGail} R. de Gail {\it et al.}, Phys. Rev. B {\bf 84}, 045436 (2011).

\bibitem{choi} M-Y Choi, H-Y Hyun and Y. Kim, Phys. Rev. B {\bf 84}, 195437 (2011).

\bibitem{freitag} M. Freitag Nature Physics {\bf 7},596 (2011).

\bibitem{millie} M.S. Dresselhaus and G. Dresselhaus, Advances in Physics {\bf 51}, 1 (2002) (p. 69).

\bibitem{offset} Because of the rotation of $\sigma_\theta$ in Eqn. 6 this point degeneracy is slightly shifted along the line
that bisects $\Delta \vec K$. This does not significantly change the physics.

\bibitem{height} E.Y. Andrei (unpublished).



\end{thebibliography}
\end{document}